\documentclass[10pt,twoside,dvips]{IEEEtran}
\usepackage{graphicx}
\input{epsf}
\newcommand{\beq}{\begin{equation}}
\newcommand{\eeq}{\end{equation}}

\begin{document}
\title{Free-Space Antenna Field/Pattern Retrieval\\ 
in Reverberation Environments}
\author{Vincenzo Fiumara, Adele Fusco, Vincenzo Matta, Innocenzo M. Pinto
\thanks{V. Fiumara and V. Matta are with $D.I^3.E$, University of Salerno, Italy;
A. Fusco and I.M. Pinto are with The Waves Group, University of Sannio at Benevento, Italy.
E-mail: pinto@sa.infn.it}}
\maketitle
\begin{abstract}
Simple algorithms for retrieving free-space antenna field or directivity patterns from complex (field) or real (intensity) measurements taken in ideal reverberation environments are introduced and discussed.
\end{abstract}
\markboth{V. Fiumara et al.}{Free-Space Radiation Field/Pattern Retrieval...}
\section{Introduction} 
\label{sec:intro}
Antenna measurements are usually performed
under simulated free-space conditions, 
e.g. by placing the antenna under test (henceforth AUT)
as well as the measuring probe in an open test-range
or in an electromagnetic (henceforth EM)
anechoic chamber \cite{Ante_Meas}.

Retrieving free-space antenna field and/or directivity patterns 
from measurements taken in any {\em realistic} (i.e., imperfect)
test-range or anechoic chamber
relies on the possibility of reconstructing
the ray skeleton of the measured field 
using robust spectral estimation
techniques, including, e.g., periodogram,
Prony,  Pisarenko, Matrix-Pencil and Gabor  algorithms \cite{SA1}-\cite{SA4},
so as to "subtract" all environment-related reflected/diffracted fields.

The direct (free-space) field, however,
{\em cannot} be unambiguously identified, unless
additional assumptions are made about its 
relative intensity and/or phase,
which {\em do not} hold true in the most general case.

A possible way to uniquely extract
the free-space (direct path) field
is to average over many measurements obtained 
by suitably changing the position of the source-probe pair
with respect to the environment, while keeping
the source-probe mutual distance and orientation fixed.
This obviously leaves the
free-space direct-path term unchanged, while
affecting both the amplitudes and the phases 
of all environment-related reflected/diffracted fields.
In the limit of a large number of measurements,
one might expect that these latter
would eventually average
to zero.
This is the rationale behind the idea
of retrieving free-space antenna parameters 
from measurements taken in
a  {\em reverberation enclosure} (henceforth RE),
where the chamber boundary is effectively moved through
several positions by mechanical stirring,
while the source-probe distance
and mutual orientation is fixed.

Through the last decades reverberation enclosures 
earned the status of elicited EMI-EMC assessment tools \cite{RE_tut}.
On the other hand, only recently
effective procedures 
for estimating antenna parameters, including
efficiency \cite{Kil_eff}, 
diversity-gain \cite{Kil_DG},
MIMO-array channel capacity \cite{Kil_MIMO},
and free-space radiation resistance \cite{Kil_RR},
from measurements made in a reverberation chamber
have been introduced by Kildal and co-workers,
in a series of pioneering papers. 

Here we discuss, perhaps for the first time,
free-space antenna  field/directivity pattern
retrieval from measurements taken 
in a reverberation environment.

The paper is organized as follows. 
In Sect. \ref{sec:revenv}
the key relevant properties of reverberation enclosure fields are summarized.
In Sect. \ref{sec:cmplx} and \ref{sec:real} 
simple algorithms for retrieving free-space
antenna field or directivity patterns, respectively 
from (complex) field or (real) intensity measurements made
in a reverberation environment are discussed,
including the related absolute and relative errors.
The related efficiency 
is the subject of Sect. \ref{sec:acc_eff}, including
some useful concepts on Cramer-Rao bounds.
Conclusions follow under Sect. \ref{sec:conclu}. 

\section{Fields in Reverberation Environments}
\label{sec:revenv}

In the following we shall sketch and evaluate some
straightforward procedures 
to estimate the free-space antenna field 
or directivity pattern from measurements
made in a reverberation enclosure.

The AUT  field/intensity
will be sampled at a suitable number of points $P$ 
of the  AUT-centered sphere $r\!=\!R$, corresponding
to as many sampling directions.
At {\em each} point $P$ we shall actually make $N$ measurements
in the reverberation environment
corresponding to as many {\em different positions} of the 
mode stirrers.

Throughout the rest of this paper 
we shall restrict to the simplest case where both the antenna
under test and the field-probe
(henceforth FP) are linearly (co)polarized,
and placed in an ideal (fully-stirred) reverberation
environment.

The relevant component of the complex electromagnetic
(henceforth EM) field at a point $P$ can be written:
\beq
E(P,n)\!=\!E_{d}(P)\!+\!E_{r}(P,n),~~n\!=\!1,2,\!\dots\!,N.
\label{eq:ensemble}
\eeq
The first term in (\ref{eq:ensemble}) 
is the {\em direct} field, 
and is the only term which would exist in free-space;
the second term is the (pure) reverberation field,
whose value depends on the stirrers' positions\footnote{
We consistently include in the free-space
antenna-field  {\em any} reflected/diffracted term  
which {\em does not} change 
as the positions of the mode stirrers change over.},
and $n$ identifies the different stirrers' positions.

According to a widely accepted model  \cite{Hill}, 
for any fixed $P$, the set $\{ E_{r}(P,n)|n=1,2,\dots,N \}$, 
can be regarded as an
{\em ensemble} of  identically distributed 
(pseudo) {\em random variables} 
resulting from the superposition 
of a large number of plane waves with uniformly
distributed phases and arrival directions.

Under these ideal (but {\em not} unrealistic) assumptions
the real and imaginary part of the reverberation field
$E_{r}(P,n)$ will be gaussian distributed\footnote{
In this connection, the amplitude distribution 
of the contributing plane waves
turns out to be almost irrelevant \cite{AliBaba}.}
and uncorrelated, with zero averages and equal variances \cite{Kost_Bov},
\beq
\langle \mbox{Re}^2 E_{r}(P,n) \rangle=
\langle \mbox{Im}^2 E_{r}(P,n) \rangle=
\frac{E_0^2}{2},
\label{eq:vars}
\eeq
where $\langle\cdot\rangle$ denotes, 
more or less obviously, statistical averaging.
The quantity $E_0^2$ in (\ref{eq:vars}) is given by \cite{Hill}:
\beq
E_0^2=\frac{8\pi\eta_0}{\lambda^2}~\Pi_r,
\label{eq:E0}
\eeq
where $\eta_0$ is the free-space wave impedance,
$\lambda$ the wavelength, and $\Pi_r$ the power received 
by {\em any} (linearly polarized, matched) antenna 
placed in the reverberation enclosure, irrespective
of its orientation and directivity diagram \cite{Hill}.
This latter is related to the total power $\Pi_t$
radiated into the enclosure by the AUT as follows \cite{CorLatPao} :
\beq
\Pi_r=\Xi~\Pi_t,
\label{eq:reverb_fact}
\eeq
where the (frequency dependent) RE 
calibration-parameter $\Xi$
is related to the chamber (internal) surface $\Sigma$
and wavelength $\lambda$ by \cite{CorLatPao}
\beq
\Xi = \frac{\lambda^2}{2 \alpha \Sigma},
\label{eq:Xi}
\eeq
$\alpha$ being an average-equivalent 
wall absorption coefficient\footnote{
The coefficient $\alpha$ in (\ref{eq:Xi})
can be evaluated as
$\alpha=\Sigma_0/(\Sigma+\Sigma_0) < 1$,
$\Sigma_0$ being the area of a wall aperture  
which halves $\Pi_r$ \cite{CorLatPao}.}.

\section{AUT Free-Space Field Estimator} 
\label{sec:cmplx}

Under the made assumption where the real and imaginary
part of the reverberation field $E_{r}$
are independent, zero-average gaussian random variables,
it is natural to adopt the following estimator of 
the free-space (complex) AUT field at $P$
in terms of the (complex) fields (\ref{eq:ensemble}):
\beq
\widehat{E}_d(P)=
N^{-1}\sum_{n=1}^{N} 
E(P,n).
\label{eq:cmplx}
\eeq
Equation (\ref{eq:cmplx}) provides unbiased estimators of 
$\mbox{Re}[E_d(P)]$ and $\mbox{Im}[E_d(P)]$, with variances
\beq
VAR[\mbox{Re}\widehat{E}_d(P)]=
VAR[\mbox{Im}\widehat{E}_d(P)]=
\frac{E^2_0}{2N}.
\label{eq:var_field}
\eeq
The related absolute and relative errors are:
\beq
\epsilon_{abs}^{(F)}=
\left\langle
\left|
\widehat{E}_d(P)-E_d(P)
\right|^2
\right\rangle^{1/2}=
N^{-1/2} E_0,
\label{eq:field_var}
\eeq
and
\beq
\epsilon_{rel}^{(F)}=
\frac{
\left\langle
\left|
\widehat{E}_d(P)-E_d(P)
\right|^2
\right\rangle^{1/2}
}{\left|E_d(P)\right|}=
\left(\frac{2}{N\theta}\right)^{1/2},
\label{eq:epsilonprime}
\eeq
where, for later convenience, we introduced the dimensionless quantity
\beq
\theta(P)=\frac{2|E_d(P)|^2}{E_0^2}.
\label{eq:distrib_int}
\eeq

The  r.m.s.  absolute 
error (\ref{eq:field_var})
can be made  as small as one wishes, in principle,
by increasing $N$, and/or the chamber size
(the distance between the chamber walls and the AUT-FP pair),
which makes $E_0 \propto \Sigma^{-1/2}$ smaller.
Keeping the AUT-FP pair distance fixed,
this will at the same time make $\theta$ larger,
in view of eq. (\ref{eq:distrib_int}), thus reducing 
the relative error (\ref{eq:epsilonprime}), when meaningful, as well.
Note that this is true for both far and near-field measurements.

\section{AUT Directivity Estimator}
\label{sec:real} 

The AUT directivity can be estimated from
(far field) intensity measurements  
made in a reverberation enclosure as follows.
Let 
\beq
I(P,n)\!=\!
\mbox{Re}^2 E(P,n)
\!+\!
\mbox{Im}^2 E(P,n),~~
n\!=\!1,2,\!\dots\!,N.
\label{eq:intensity}
\eeq
It is convenient to scale the field intensities $I(P,n)$ 
to the variance in (\ref{eq:vars}), by
letting $\xi_n(P)=2I(P,n)/E_0^2$, so that all the $\xi_n$ 
are (identically) distributed
according to a {\em noncentral} chi-square 
with two degrees of freedom \cite{Kay}
and non-centrality parameter  $\theta(P)$  given by eq. (\ref{eq:distrib_int}).

We may use the obvious far field formula:
\beq
|E_d(P)|^2=\frac{2\eta_0 \Pi_t}{4\pi R^2}~D(P),
\eeq
where $D(P)$ is the AUT directivity,
together with eq.s (\ref{eq:E0}) and (\ref{eq:reverb_fact}) in
eq. (\ref{eq:distrib_int}) to relate $\theta$ to the AUT directivity $D$ as follows
\beq
\theta=\frac{1}{8\pi^2\Xi(f)}\left(\frac{\lambda}{R}\right)^2 D =: \gamma D,
\label{eq:theta}
\eeq
where the dependence of $D$ and $\theta$
on the measurement point (direction) is understood 
and dropped for notational ease\footnote{
For the simplest case of a spherical enclosure of radius $R_c$, 
from eq.s (\ref{eq:Xi}) and (\ref{eq:theta}) one gets
$\theta=(\alpha/\pi)(R_c/R)^2 D$, $R$ being the AUT-FP distance.}.
The probability density function of the $\xi_n$
can be accordingly written \cite{Kay} 
\beq
f(\xi;D)=\frac{1}{2} \exp
\left(-\frac{\xi+\gamma D}{2}\right)
I_0(\sqrt{\xi\gamma D}).
\label{eq:CDF_xi}
\eeq
The first two moments accordingly are:
\beq
\langle \xi \rangle=2+\gamma D,~~~VAR[\xi]=4+4\gamma D,
\label{eq:2mom}
\eeq
which suggest using the following (simplest, unbiased) estimator of $D$ \cite{Johnson}:
\beq
\widehat D=
(\gamma N)^{-1}
\sum_{n=1}^N (\xi_n-2),
\label{eq:int_est}
\eeq
for which
\beq
VAR[\widehat D]=
\frac{4}{\gamma^2 N}\left(1+\gamma D\right).
\label{eq:int_est_var}
\eeq
The absolute and relative errors of the directivity estimator (\ref{eq:int_est}) are thus:
\beq
\epsilon_{abs}^{(D)}=\sqrt{VAR[\widehat D]}
\label{eq:abs_err_D}
\eeq
and:
\beq
\epsilon_{rel}^{(D)}\!=\!
D^{-1}
\sqrt{
\left\langle
\!
\left(
\widehat{D}\!-\!D
\right)^2
\right\rangle
}\!=\!
\frac{2}{\theta}\left(\frac{1+\theta}{N} \right)^{1/2}.
\label{eq:RMS_ERR}
\eeq

The  absolute and  relative (when meaningful)
errors (\ref{eq:abs_err_D}) and  (\ref{eq:RMS_ERR})
can  be made  as small as one wishes, in principle,
by increasing $N$, and/or the chamber size,
so as to make $\theta$ suitably large, in view of (\ref{eq:theta}).

Note that in all derivations above 
we made the implicit assumption of dealing with  
{\em independent} measurements.

The number $N$ of independent measurements
needed to achieve relative errors $\sim 5 \cdot 10^{-2}$ 
for both field and directivity measurements
is shown in Fig. 1,
\begin{figure}[htbp]
	\begin{center}
		\includegraphics[width=7cm,height=7cm]{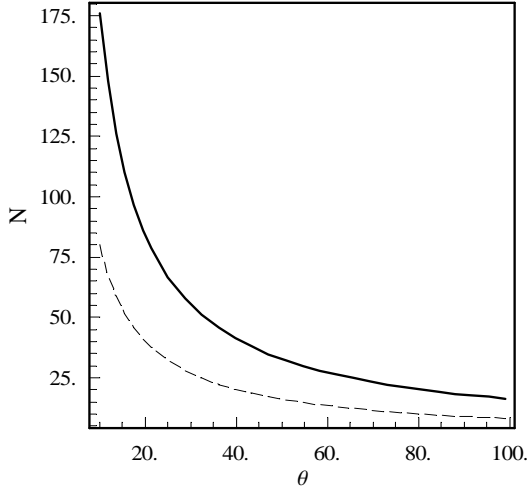}
  	\end{center}
	\label{fig:numsam}
	\caption{Number $N$ of independent measurements vs. $\theta$ needed 
	for relative errors $\sim 5\cdot 10^{-2}$.
	Solid line: $\epsilon_{rel}^{(D)}$; dashed line: $\epsilon_{rel}^{(F)}$.}
\end{figure}
and is of the order of $10^2$ for $\theta \sim 10$.
This figure is consistent with typical experimental findings \cite{samples},
and also with theoretical estimates obtained
from a chaos-based models of reverberation enclosures \cite{chaos}.

\section{Efficiency of Proposed Estimators}
\label{sec:acc_eff}

An obvious question is now whether one could do better
using {\em different} estimators, other than 
(\ref{eq:cmplx}) and (\ref{eq:int_est}).

The natural benchmark for gauging the goodness
of an estimator is the well-known Cramer-Rao lower bound (henceforth CRLB) \cite{Cramer}.
We limit ourselves here to remind a few basic
definitions and properties.
Let ${\mathbf \xi}= \{\xi_k|k=1,2,\dots,K\}$
a set of (real) random variables with joint probability density
$f({\mathbf \xi};{\mathbf X})$, where ${\mathbf X}=\{X_m|m=1,2,\dots,M\}$ 
is a set of (unknown, real) parameters to be estimated.
One can prove that\footnote{
We implicitly assume that  
the following regularity condition \cite{Kay} holds:
$\left\langle
\frac{\partial \log f({\mathbf \xi},{\mathbf X})}{\partial {\mathbf X}}
\right\rangle=0$.}
for  any estimator 
$\widehat{{\mathbf X}}$ of ${\mathbf X}$
such that 
$\langle\widehat{{\mathbf X}}\rangle = {\mathbf X}$,
({\em unbiased} estimator), one has:
\beq
{\mathbf C}_{\widehat{\mathbf X}}-
{\mathbf J}^{-1}({\mathbf X})\geq 0
{\label{eq:CramerRaoVec}}
\eeq
where ${\mathbf C}$ is the covariance matrix, viz.:
\beq
\left[
{\mathbf C}_{\widehat{\mathbf X}}
\right]_{hk}=
\left\langle
(\widehat{X}_h-X_h)
(\widehat{X}_k-X_k)
\right\rangle,
\label{eq:CRLBmat}
\eeq
\beq
\left[
{\mathbf J}
({\mathbf X})
\right]_{hk}=-\left\langle
\frac{\partial^2\log{f({\mathbf \xi};{\mathbf X})}}{\partial X_h\partial X_k}
\right\rangle
\label{eq:FisherMat}
\eeq
is the Fisher information matrix,
the expectations are taken with respect to $f({\mathbf \xi};{\mathbf X})$,
and the true value of ${\mathbf X}$ is used for evaluating
(\ref{eq:FisherMat}).
Equation (\ref{eq:CramerRaoVec}) implies $M$ inequalities
whereby the diagonal elements of 
${\mathbf C}_{\widehat{\mathbf X}}$, i.e.,
the variances of the  components of  $\widehat{{\mathbf X}}$,
are bounded from below. These are the CRLB s.
An estimator for which the l.h.s. of 
eq. (\ref {eq:CramerRaoVec}) is actually zero, i.e., for which
the variance of each  component of  $\widehat{{\mathbf X}}$
attains its CRLB is called {\em efficient}.
For the special case where the $\xi_k$ are independent
and identically distributed, with a PDF $f(\xi;X)$
depending on a {\em single} parameter $X$,
equation (\ref{eq:CramerRaoVec}) becomes
$$
VAR[\widehat X] \ge -
\frac{1}
{\displaystyle{N \left\langle
\frac{\partial^2 \log{f(\xi;X)}}{\partial X^2}
\right\rangle}}=
$$
\beq
=\frac{1}
{\displaystyle{N 
\left\langle
\left[
\frac{\partial \log{f(\xi;X)}}{\partial X}
\right]^2
\right\rangle}}.
\label{eq:CramerRao1}
\eeq

One can readily prove that 
the field estimator (\ref{eq:cmplx}) is an
efficient one, since the r.h.s of (\ref{eq:var_field})
coincides with the pertinent CRLB.
The simplest directivity estimator (\ref{eq:int_est}),
on the other hand, while {\em not} efficient,
gets very close to its CRLB, as shown below.

The Cramer-Rao bound for 
the estimator (\ref{eq:int_est}),
is obtained by using the following formula,
which  follows directly from eq. (\ref{eq:CDF_xi}),
\beq
\frac{\partial \log f(\xi;D)}{\partial D}=
-\frac{\gamma}{2}+\frac{1}{2}\frac{I_1\left(\sqrt{\xi\gamma D}\right)}
{I_0\left(\sqrt{\xi\gamma D}\right)}
\sqrt{\frac{\xi\gamma}{D}},
\label{eq:fisher_int}
\eeq
where $I_{0,1}(\cdot)$ are modified Bessel functions,
and is \cite{Johnson}:
\beq
CRLB(D)=\frac{4}{\gamma^2 N}
\left(
\frac{\Lambda}{\gamma D}-1
\right)^{-1},
\label{eq:pierrozzo}
\eeq
where
\beq
\Lambda=\left\langle
\xi I_1^2\left(\sqrt{\xi\gamma D}\right) \cdot I_0^{-2}\left(\sqrt{\xi \gamma D}\right)
\right\rangle,
\label{eq:CRLB}
\eeq
and the expectation is taken with respect to $f(\xi;D)$.

The ratio between the CRLB (\ref{eq:CRLB}) and the variance (\ref{eq:int_est_var})
yields the {\em relative efficiency} 
\beq
\rho=\frac{CRLB(D)}{\mbox{VAR}[\widehat{D}]}.
\label{eq:rel_eff}
\eeq
The relative efficiency (\ref{eq:rel_eff}) 
of the proposed directivity estimator (\ref{eq:int_est})
is readily computed from eq.s 
(\ref{eq:pierrozzo}), (\ref{eq:CRLB}) 
and (\ref{eq:int_est_var}), 
and is independent of $N$.
It is displayed  in Fig. 2 vs. $\theta=\gamma D$.
\begin{figure}[htbp]
	\begin{center}
		\includegraphics[width=7cm,height=7cm]{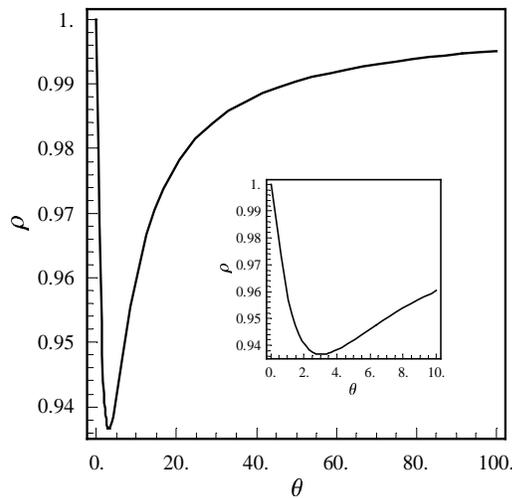}
	\end{center}
	\label{fig:effic}
	\caption{– Directivity estimator. Relative efficiency $\rho$,
	 eq. (\ref{eq:rel_eff}), vs. $\theta$, eq. (\ref{eq:theta}).}
\end{figure}
The relative efficiency of (\ref{eq:int_est})
is seen to be pretty decent, being always larger than $\approx .937$.

\section{Conclusions}
\label{sec:conclu} 

Free-space antenna field/directivity measurements 
in ideal reverberation enclosures
have been shortly described and evaluated.
The main simplifying assumptions (linearly co-polarized 
AUT and FP) can be more or less easily relaxed
at the expense of minor formal complications which do
not alter the main conclusions.
On the basis of these preliminary results, 
the possibility of performing cheap, simple and reliable 
{\em in situ} antenna measurements using, e.g., flexible
conductive thin-film deployable/inflatable enclosures with air-blow stirring  
\cite{greek}, \cite{VIRC} is envisaged.



\end{document}